\documentclass[preprint,aps]{revtex4}
\usepackage{color}
\usepackage{mathtools}
\usepackage{array}
\usepackage{tabularx}
\usepackage{diagbox}
\usepackage{tikz}
\usetikzlibrary{decorations.pathmorphing}
\usepackage{bm}
\usepackage{amsmath,bm,amssymb,amsfonts,dcolumn,color,graphicx,latexsym,epsfig}
\usepackage{rsfso}
\usepackage{hyperref}
\hypersetup{
colorlinks=true,
linkcolor=magenta,
filecolor=magenta,      
urlcolor=blue,
citecolor=blue,
}
\usepackage{multirow}
\usepackage{natbib}
\usepackage{float}
\usepackage{booktabs}
\usepackage{pifont}
\usepackage{mathrsfs}
\usepackage{caption}
\usepackage{caption, threeparttable}
\begin{document}

\title{Pulse-induced memory-like effect in cyclotron motion?}
\author{Sayan Kar}
\email{sayan@phy.iitkgp.ac.in}
\affiliation{Department of Physics, Indian Institute of Technology, Kharagpur 721 302, India}

\begin{abstract}
\noindent We study how a charged particle moving in a uniform magnetic field
along its standard circular path (cyclotron motion) reacts to a short-duration, homogeneous, uniform electric field pulse
injected in the plane perpendicular to the magnetic field. A
`permanent' change
in the radius of the initial circle and a shift of its centre is noted at
later times, after the pulse is switched off. The velocity vector (components and magnitude)
undergoes a change too, akin to a `velocity kick'. The cause behind this 
permanent change appears to be linked to the difference 
between the vector potentials before and
after the duration of the electric pulse. Further, we show how such vector potentials in the past and future are related 
through a gauge transformation. In summary, our
results suggest a pulse-induced `electromagnetic memory-like effect', which is not quite a
`wave memory', but, nevertheless, has similar features within a 
simple, non-relativistic context.

\noindent 
\end{abstract}


\maketitle

\newpage

\section{Introduction}

\noindent Gravitational wave memory--a `permanent change' caused by a gravitational wave pulse--has been a topic of active research interest
over the last few years. Though known to us since the papers of
Zel'dovich, Polnarev \cite{zelpol} and Braginsky, Grishchuk \cite{bragri} 
and subsequent seminal work by many authors (eg. \cite{christo}),
the recent upsurge in research is largely motivated by prospects of
observations in the context of gravitational wave physics \cite{memoryexpt0, memoryexpt1,memoryexpt2}. Different aspects of gravitational wave memory
have been analysed in a large number of papers \cite{biegarf,tolwal,tolwal1,madwin1,madwin2,stro1,stro2,zhang1,zhang2,zhang3,zhang4,zhang5,ic1, cvet,ic2,ic3,ic4,pah1,pah2,jpuzan,eanna1,eanna2,eanna3,olough,diva,shore,mota}.
Reviews on gravitational wave memory are
available in \cite{favata1}, \cite{cqgfocus1}.

\noindent  In analogy with gravitational wave memory, electromagnetic memory
is conventionally defined also through a `permanent change'-- the so-called `velocity kick'. It was first briefly discussed in a paper by Grishchuk and Polnarev \cite{grishpol} and has recently been extensively analysed by Bieri and Garfinkle \cite{biegarf1}. It is defined (for a specific case where the
force is $q {\bf E}$ and unit mass) using the following simple equation:
\begin{eqnarray}
{\bf v}_{\infty} - {\bf v}_{-\infty} = q \int_{-\infty}^{\infty} {\bf E}\, dt
\end{eqnarray}
where ${\bf E}$ is the electric field, usually taken as a radiation field due to
a source far away. In the above formula, if one considers an electric field which is non-zero
and constant only over a small interval (0,T) in time, then the R. H. S.
simply becomes ${q\bf E_0} T$ and we have the velocity at negative infinity (past)
having a different value compared to its value at positive infinity (future). This also 
implies a jump or a kick in the velocity, imparted to the charged test particle
by the electric pulse. Conventionally though, one considers the electric field due to
radiation--say,
electric dipole radiation (far away from the source) -- and its effect on a test charge. In the non-relativistic case, one can evaluate the time rate of change of the dipole moment of the source in the past and future and relate it to the charges and velocities of the source. This would give an expression for the velocity kick on the test charge in terms of the behaviour of the source in the past and future \cite{biegarf1}. The analysis is somewhat more intricate when one moves away from the
slow motion approximation for the source entities. In such a fully relativistic
scenario one uses the outgoing null coordinate $u$ instead of time $t$ 
and evaluations are done at future (past) null infinity 
($r\rightarrow\infty$) using $u$. The discussion is available in
\cite{biegarf1} and also nicely summarised briefly in \cite{jokela}.
The essence however is to see if the velocity of the test charge changes on account of the field or the pulse (i.e. we find a nonzero value of the L. H. S. in Eqn. (1)), and whether there is a
`permanent change' which appears after the pulse has departed, and persists later too. It is this viewpoint, which does not necessarily attribute memory to a passing wave field at null infinity, is what we focus on in this article. In a sense, our approach to `memory' as an observable phenomenon is somewhat similar to the idea of `persistent observables'
extensively pursued in \cite{eanna1,eanna2,eanna3}.
Diverse aspects of electromagnetic memory as well as `color' (Yang-Mills) memory have been discussed in numerous papers in the recent past
\cite{winicour,susskind, pasterski,pate,satish,andrea,garfinkle,biegarf2,jokela, seraj}.

\noindent In the example which we discuss here, we will consider a
particle of charge $q$ and mass $m$ placed in a uniform magnetic field and encountering
a spatially  homogeneous electric pulse. Prior to the arrival of the pulse, the particle travels in a circular path.
After the pulse has departed, what happens to this trajectory? How is it
influenced by the pulse? Is that influence a permanent change? These
questions when answered (as we do below) will shed light on an effect 
quite similar
to `electromagnetic memory' , {\em albeit} in a rather simple, purely non-relativistic context. We emphasize 
here, that we are {\em not} considering a `wave' memory--rather we are defining `memory' as a remembrance arising out of a short duration electric pulse in a background
uniform magnetic field. From our perspective, the memory aspect is largely encoded in a `persistent and permanent
change' caused by the pulse and we are interested in illustrating and quantifying this change
through our work.

\section{Charged particle in a uniform magnetic field and an electric pulse}

\noindent Let us now state our problem more quantitatively and obtain a
solution. We shall assume the magnetic field to be uniform and along the
$z$ direction, i.e. ${\bf B} =B {\bf \hat k}$. The electric field is in the
$xy$ plane and we assume it to be:
\begin{eqnarray}
    {\bf E} = E (t) \left (\cos \,\alpha {\bf \hat i} + \sin \, \alpha {\bf \hat j}
    \right )
\end{eqnarray}
where $E$ is chosen as the following function (see Figure 1):
\begin{eqnarray}
E(t) = 0, \hspace{0.4in} -\infty \le t \leq 0
\\ \nonumber
= E, \hspace{0.5in} 0\leq t \leq T \\ \nonumber
= 0 \hspace{0.9in} t \ge T
\end{eqnarray}
with the `$E$' introduced above, a constant.
\begin{figure}[h]
\includegraphics[width=0.6\textwidth]{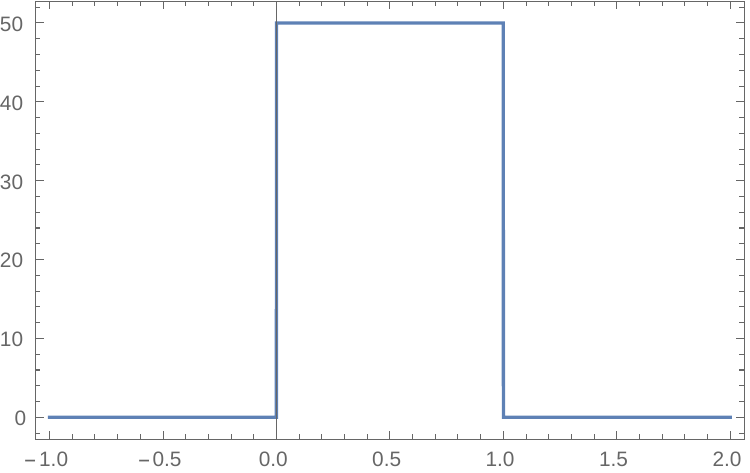}
\caption{The square pulse $E$ as function of $t$.}
\end{figure}
\noindent This is the well-known square pulse, as shown in Figure 1.

\noindent In the standard cyclotron, the electric field is chosen to be an alternating harmonic field
applied between the dees and oscillating at a frequency in the radio-frequency range. This leads to a spiral path 
and the charged particle gets accelerated in the process, thereby achieving the aim of
an accelerator \cite{chautard}. In contrast, here we apply a short duration electric field with the aim 
of injecting a visible difference between trajectories before and after the duration of the pulse.

\noindent The equations of motion for our system in the interval $(0,T)$ (where the electric pulse acts) are easy to write down ($m \frac{d{\bf v}}{dt} = q {\bf E} + q {\bf v} \times {\bf B}$) and are given, in component form, as:
\begin{eqnarray}
m\frac{dv_x}{dt} = q E \cos \alpha + q v_y B \\
m \frac{dv_y}{dt} = q E \sin \alpha - q v_x B \\
m \frac{dv_z}{dt} = 0
\end{eqnarray}
where $q$ and $m$ are the charge and mass of the particle, respectively, and ${\bf v} = \left (v_x,v_y, v_z\right )$ is its velocity vector.
In the other two regions, i.e. for $t\leq 0$ and $t \geq T$, the same equations 
hold with $E=0$. The $v_z$ equation is trivially solved and we shall assume
$v_z=0$ by setting the integration constant to zero, thereby restricting the
motion of the charge to the $xy$ plane.

\noindent The well-known way to solve the equations for $v_x$ and $v_y$ is to
differentiate one of them first and use the other equation. This will give a
second order uncoupled equation for $v_x$ (or $v_y$). One can then use the
solution of this second order equation to find the other velocity component
using one of the first order equations. 

\noindent For the problem at hand, we will need to write down solutions
for $x (t)$, $y(t)$, $v_x(t)$, $v_y(t)$ in each of the three regions:
I ($t\leq 0$), II($0\leq t \leq T$ and III($t\geq T$). The solutions for
the coordinates and velocities have to be matched at the two boundaries
at $t=0$ and $t=T$. This problem appears to have been discussed briefly in
a very old paper \cite{gould}, though the analysis there
addresses a different motivation and purpose. 

\noindent Going through this rather straightforward exercise, we
obtain all solutions in the three regions. These are given below.
Here, $\omega_B = \frac{q B}{m}$ is the so--called cyclotron frequency, and the superscripts `I', `II' and `III'
refer to the region under consideration, as mentioned above. Arbitrary
integration constants appear which will be fixed through 
chosen conditions on the coordinates and velocities. Note that there are four constants which can be
fixed through conditions on position and velocity. We will, for now, keep them unspecified in the expressions below. While discussing examples and 
showing plots
later, we will choose representative values.

\noindent {\bf Region I ($t\leq 0$):}
\begin{eqnarray}
x^{(I)} = A_0 \sin \, \omega_B t- D_0 \cos \, \omega_B t + C_0 \\
y^{(I)} = A_0 \cos \, \omega_B t + D_0 \sin \, \omega_B t + C'_0 
\end{eqnarray}
\begin{eqnarray}
v_{x}^{(I)} = \omega_B A_0 \cos \, \omega_B t + \omega_B D_0 \sin \, \omega_B t \\
v_{y}^{(I)} = -\omega_B A_0 \sin \, \omega_B t + \omega_B D_0 \cos \, \omega_B t
\end{eqnarray}
\noindent {\bf Region II ($0\leq t \leq T$):}
\begin{eqnarray}
x^{(II)} = \left ( \frac{E}{B}\sin \, \alpha\right )t + \left (A_0 - \frac{E}{\omega_B B} \sin \, \alpha \right ) \sin \, \omega_B t \nonumber \\ - \left (D_0 + \frac{E}{\omega_B B}\cos \, \alpha \right ) \cos \, \omega_B t  + \left ( C_0 +\frac{E}{\omega_B B}
\cos \, \alpha \right ) \\
y^{(II)} = -\left ( \frac{E}{B}\cos \, \alpha\right )t + \left (A_0 - \frac{E}{\omega_B B} \sin \, \alpha \right ) \cos\, \omega_B t \nonumber \\ + \left (D_0 + \frac{E}{\omega_B B}\cos \, \alpha \right ) \sin \, \omega_B t  + \left ( C'_0 +\frac{E}{\omega_B B}
\sin \, \alpha \right ) 
\end{eqnarray}
\begin{eqnarray}
v_{x}^{(II)} = \left ( \frac{E}{B}\sin \, \alpha\right ) + \omega_B \left (A_0 - \frac{E}{\omega_B B} \sin \, \alpha \right ) \cos \, \omega_B t \nonumber \\ +\omega_B \left ( D_0 + \frac{E}{\omega_B B}\cos \, \alpha \right ) \sin \, \omega_B t \\
v_{y}^{(II)} = -\left ( \frac{E}{B}\cos \, \alpha\right ) - \omega_B \left (A_0 - \frac{E}{\omega_B B} \sin \, \alpha \right ) \sin \, \omega_B t \nonumber \\ +\omega_B \left ( D_0 + \frac{E}{\omega_B B}\cos \, \alpha \right ) \cos \, \omega_B t
\end{eqnarray}
\noindent {\bf Region III ($t\geq T$):}
\begin{eqnarray}
x^{(III)} = \left [ A_0 +\frac{2 E}{\omega_B B }\sin \, \frac{\omega_B T}{2} \cos \left (\frac{\omega_B T}{2} +\alpha \right ) \right ] \sin \, \omega_B t \nonumber \\ -\left [ D_0 +\frac{2 E}{\omega_B B }\sin \, \frac{\omega_B T}{2} \sin \left (\frac{\omega_B T}{2} +\alpha \right ) \right ] \cos \, \omega_B t
\nonumber \\ + \left ( C_0 + \frac{E}{B} \left (\sin \, \alpha \right ) T \right ) \\
y^{(III)} = \left [ A_0 +\frac{2 E}{\omega_B B }\sin \, \frac{\omega_B T}{2} \cos \left (\frac{\omega_B T}{2} +\alpha \right ) \right ] \cos \, \omega_B t \nonumber \\ +\left [ D_0 +\frac{2 E}{\omega_B B }\sin \, \frac{\omega_B T}{2} \sin \left (\frac{\omega_B T}{2} +\alpha \right ) \right ] \sin \, \omega_B t 
\nonumber \\ + \left ( C'_0 - \frac{E}{B}\left (\cos \, \alpha\right ) T \right ) 
\end{eqnarray}
\begin{eqnarray}
v_{x}^{(III)} =  \omega_B\left [ A_0 +\frac{2 E}{\omega_B B }\sin \, \frac{\omega_B T}{2} \cos \left (\frac{\omega_B T}{2} +\alpha \right ) \right ] \cos \, \omega_B t \nonumber \\ +\omega_B \left [ D_0 +\frac{2 E}{\omega_B B }\sin \, \frac{\omega_B T}{2} \sin \left (\frac{\omega_B T}{2} +\alpha \right ) \right ] \sin \, \omega_B t \\
v_{y}^{(III)} = -\omega_B \left [ A_0 +\frac{2 E}{\omega_B B }\sin \, \frac{\omega_B T}{2} \cos \left (\frac{\omega_B T}{2} +\alpha \right ) \right ] \sin \, \omega_B t \nonumber \\ +\omega_B \left [ D_0 +\frac{2 E}{\omega_B B }\sin \, \frac{\omega_B T}{2} \sin \left (\frac{\omega_B T}{2} +\alpha \right ) \right ] \cos \, \omega_B t 
\end{eqnarray}
One may verify that the above coordinate solutions and velocity components match at the boundaries at $t=0$ and $t=T$. The time derivative of the velocity components are however discontinuous at $t=0$ and $t=T$ because of the
discontinuous nature of the electric pulse, a fact evident from Eqns. (3), (4) and (5).

\noindent We also know that the coupled equations for $v_x$, $v_y$ can be rewritten
using a complex velocity ${\bf \tilde w} = v_x+ i v_y$ and solved directly using
${\bf \tilde w} = {\bf \tilde p} \, e^{-i\omega_B t}$. 
The differential equation for the complex ${\bf \tilde p}$ in Region II
is,
\begin{eqnarray}
    \frac{d{\bf \tilde p}}{dt}=\frac{qE}{m} e^{i(\omega_B t +\alpha)}
\end{eqnarray}
In Regions I and III, the corresponding equations are found by just putting $E=0$.. 
Solving the differential equation for 
${\bf \tilde p}$ in each region and using the boundary matching for
${\bf \tilde p}$ at $t=0$ and $t=T$, the difference between the complex velocities in Regions I and III can be found. This is given as, 
\begin{eqnarray}
   {\bf \tilde w}_{(t>T)} - {\bf \tilde w}_{(t<0)}= -i \frac{E}{B} e^{-i \omega_B t} \left[ e^{i(\omega_B T+\alpha)} - e^{i\alpha} \right]
\end{eqnarray}
The velocity kick for the $x$ and $y$ components of ${\bf v}$ as
noted from (9), (10), (17), (18) 
matches with the real and imaginary parts of (20).  In a way, ((20) generalises Eqn. (1)
when both electric and magnetic fields (uniform) are present. One may consider $t$ to be arbitrarily large ($t\rightarrow \infty$, distant future) or 
small ($t\rightarrow-\infty$, remote past). Subsequently, the $x(t)$ and $y(t)$ can also be
found by integrating ${\bf w} = \frac{d{\bf z}}{dt} = \dot x + i\dot y$
and applying the boundary conditions. Later, we will obtain a different and perhaps more appropriate way of expressing the velocity kick, once we elaborate on the vector potential in the next section.

\begin{figure}[h]
\includegraphics[width=0.6\textwidth]{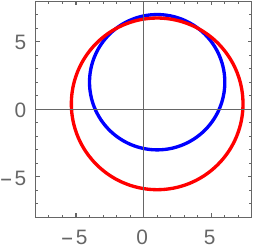}
\caption{The circular paths before and after the arrival of the pulse.
The blue (red) curve traces the motion of the charged particle before (after) the electric pulse is injected. Parameter values
chosen (in respective units) are: $A_0=3$, $D_0=4$, $C_0=1$, $C'_0=2$, $\omega_B=1$, $\alpha=2 \pi$, $T=2$, $\frac{E}{B}=0.8$. The centre of the blue curve is at $(1,2)$ and its radius is 5 units. For the red curve, the
centre is at $(1, 0.4)$ and the radius is $6.34$ units.}
\end{figure}
\begin{figure}[h]
\includegraphics[width=0.6\textwidth]{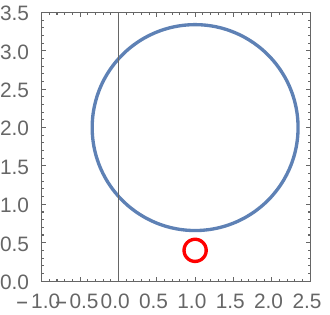}
\caption{The circular paths before and after the arrival of the pulse.
The blue (red) curve traces the motion of the charged particle before (after) the electric pulse is injected. Parameter values
chosen (in respective units) are: $A_0=-0.6$, $D_0=-1.2$, $C_0=1$, $C'_0=2$, $\omega_B=1$, $\alpha=2 \pi$, $T=2$, $\frac{E}{B}=0.8$. The centre of the blue curve is at $(1,2)$ and its radius is 1.342 units. For the red curve, the
centre is at $(1, 0.4)$ and the radius is $0.144$ units. In this case, the
radius is smaller.}
\end{figure}
\noindent Let us now turn to analysing the nature and features of the trajectories  in the different regions. In Region I, one can clearly see that we have a circular trajectory
given by the equation:
\begin{eqnarray}
\left (x^{(I)}- C_0\right )^2 + \left (y^{(I)}-C'^0\right )^2 = A_0^2 +D_0^2
\end{eqnarray}
The centre of the circle is at $(C_0,C'_0)$ and its radius is
$\sqrt{A_0^2+D_0^2}$.

\noindent The effect of the pulse is manifest in the  modified  trajectories
in regions II and III. In III, we find that the  trajectory is still a circle but with a shifted centre and
a different radius. The equation of this new circle at times $t\geq T$ is given as:
\begin{eqnarray}
\left ( x^{(III)} - \left ( C_0 + \frac{E}{B} \left (\sin \, \alpha \right ) T
\right ) \right )^2 + \left ( y^{(III)} - \left ( C'_0 - \frac{E}{B}\left (\cos \, \alpha\right ) T \right ) \right )^2 = \nonumber \\ \left [ A_0 +\frac{2 E}{\omega_B B }\sin \, \frac{\omega_B T}{2} \cos \left (\frac{\omega_B T}{2} +\alpha \right ) \right ]^2 + \left [ D_0 +\frac{2 E}{\omega_B B }\sin \, \frac{\omega_B T}{2} \sin \left (\frac{\omega_B T}{2} +\alpha \right ) \right ]^2
\end{eqnarray}

\begin{figure}[h]
\includegraphics[width=0.5\textwidth]{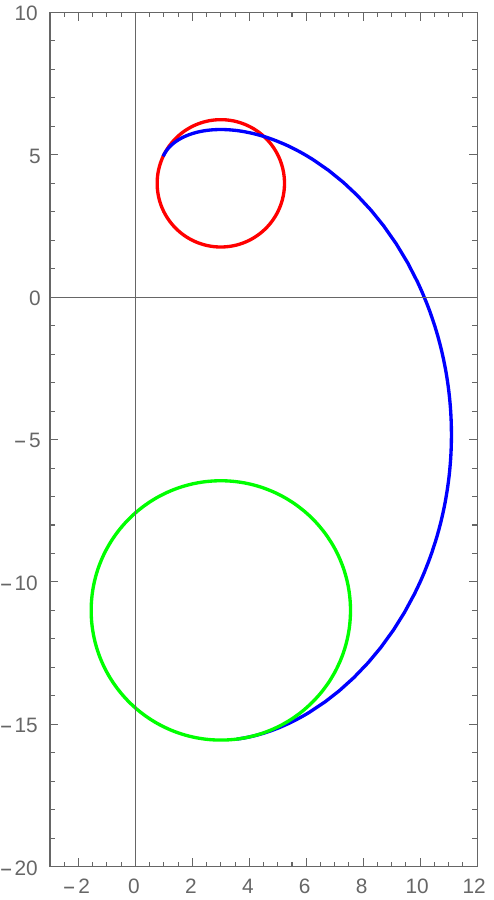}
\caption{Evolution of the charge in time. The red curve (pre-pulse circle) is from $t=-7$ to $t=0$ (clockwise). The blue curve is from $t=0$ to
$t=5$ and the green curve (post-pulse circle) is from $t=5$ to $t=15$ (clockwise). The points $t=0$ and $t=5$ are the locations where the red, blue and the blue, green curves meet tangentially, as shown. Chosen parameters are: $\omega_B=1$, $T=5$, $\alpha=2\pi$, $A_0=1$, $D_0=2$, $C_0=3$
and $C'_0=4$, $\frac{E}{\omega_B B}=3$.}
\end{figure}
\noindent Therefore, the centre of the circle now appears to be at $ \left ( C_0 + \frac{E}{B} \left (\sin \, \alpha \right )T,  C'_0 - \frac{E}{B}\left (\cos \, \alpha\right ) T \right )$. Its radius is now changed (increased/decreased) to a value equal to the 
square root of the expression 
given in the R. H. S. of Eqn. (22). It is easy to check and worth noting
that there is {\em no choice of values of any of the constants} for which both the location of the centre and the radius remain unchanged. 

\noindent The two circular paths before (blue curve) and after (red curve) the pulse are illustrated 
(using specific values of the parameters and constants) in Figures 2 and 3. The shift of the centre and  the
increase/decrease in the radius are clearly visible in both figures.

\noindent An observer watching the charge from a time much before $t=0$
will find it moving along a circular path of a certain radius. As one approaches
the time $t=0$, the charge jumps to the trajectory (not closed) as specified in Eqns. (11) and (12). When $t=T$ is reached, the trajectory joins up with a
circle with a different centre and a different radius. Thereafter, the charge is found to move on this
new circular path forever. Figure 4 shows the evolution -- the initial circle in red,
the curved path in-between in blue and the final circular path in green. The shift of the centre of the circle and the change in the radius are once again manifest here, clearly. One may extend the time to values before $t=-7$ units and beyond $t=15$ units (also note, these are just values chosen for plotting the graphs) to any extent -- in such a case we will just observe
winding around the red and green circles along which the charge moves in the remote past and distant future, respectively.

\noindent We now list a few features and consequences of the equations and results stated above. The main purpose is to point out 
specific changes caused by the pulse which highlight memory-like
characteristics. 

\

\noindent {\sf Velocities:} The change in the velocity components can be found by simply evaluating
$\Delta v_\alpha = v_\alpha(t_f)- v_\alpha (t_i)$, where $t_i<0$ and $t_f>T$ and $\alpha\equiv x,y$. For no
pair of values of $t_i$ and $t_f$ is the difference $\Delta v_\alpha =0$.
We can stretch $t_i$ to negative infinity and $t_f$ to positive 
infinity, thereby reaching asymptotic past and future.
The nonzero $\Delta v_\alpha$ ensures
the presence of a {\bf memory-like effect}. We will understand this aspect
in more detail in the next section.

\noindent One can also evaluate $v^2$ at any time $t_i<0$ and $t_f>0$.  The difference
$v_{f}^2 - v_{i}^2$ turns out  to be
\begin{eqnarray}
v_f^2-v_i^2 =  \omega_B^2 \left \{ \left [ A_0 +\frac{2 E}{\omega_B B }\sin \, \frac{\omega_B T}{2} \cos \left (\frac{\omega_B T}{2} +\alpha \right ) \right ]^2 \right . \nonumber \\ \left .  + \left [ D_0 +\frac{2 E}{\omega_B B }\sin \, \frac{\omega_B T}{2} \sin \left (\frac{\omega_B T}{2} +\alpha \right ) \right ]^2 - A_0^2 - D_0^2 \right \} 
\end{eqnarray}
This, once again shows up as a {\bf memory-like effect} (similar, but not exactly the same
as the so-called `velocity kick' \cite{biegarf1}) induced by the electric field pulse.

\noindent The velocity fields before and after the pulse have ${\bf \nabla \cdot v = 0}$
and ${\bf \nabla \times v \neq 0}$. Hence, one would say that the gradient
of the velocity field has zero expansion and shear but a nonzero rotation both before and after the electric pulse acts. The rotation is just proportional to
$\omega_B$. One cannot therefore derive any link to a memory effect from the
kinematics of a family of trajectories \cite{skrsadg}. In other words there is no
${\cal B}$ memory \cite{olough}, \cite{ic1} to be seen here.

\begin{figure}[h]
\includegraphics[width=0.6\textwidth]{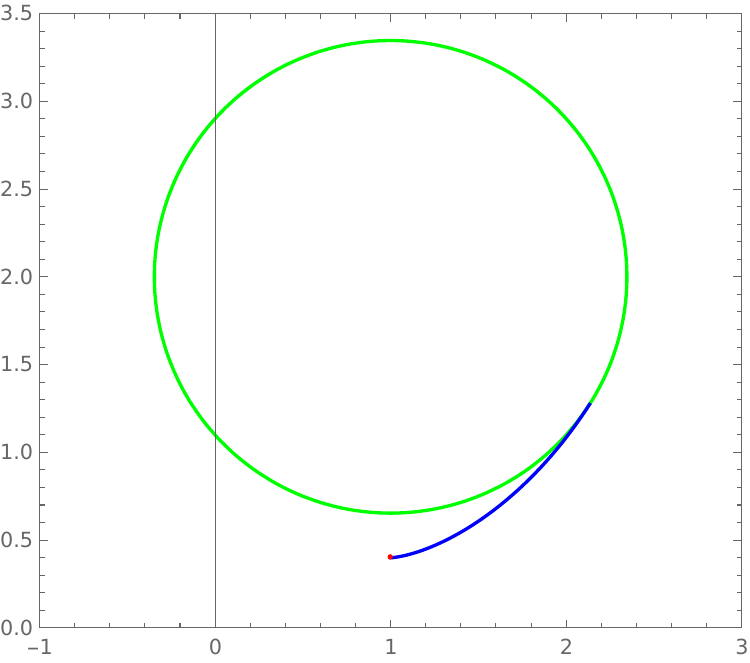}
\caption{The circular path (green) before the arrival of the pulse.
The blue curve traces the motion of the charged particle
during the period when the electric pulse is present. The red dot 
at the end of the blue curve shows the static particle after the pulse has departed. Parameter values
chosen (in respective units) are: $A_0=-0.727$, $D_0=-1.133$, $C_0=1$, $C'_0=2$, $\omega_B=1$, $\alpha=2 \pi$, $T=2$, $\frac{E}{B}=0.8$. The centre of the green curve is at $(1,2)$ and its radius is 1.342 units.
The location (red dot) of the static particle at late times  is at $(1, 0.4)$.}
\end{figure}

\

\noindent {\sf Special cases:} With properly tuned initial conditions one may end up with $x,y$ 
as constant. This will happen when the values of $A_0$ and $D_0$ are chosen to be numerically given by:
\begin{eqnarray}
A_0 = -\frac{2 E}{\omega_B B }\sin \, \frac{\omega_B T}{2} \cos \left (\frac{\omega_B T}{2} +\alpha\right ) \\
D_0 = -\frac{2 E}{\omega_B B }\sin \, \frac{\omega_B T}{2} \sin \left (\frac{\omega_B T}{2} +\alpha\right )
\end{eqnarray}
As an example, for the parameter values $C_0=1$, $C'_0=2$, $\omega_B=1$, $\alpha=2 \pi$, $T=2$, $\frac{E}{B}=0.8$ (used to generate the curves in Figure 2), the
choice of $A_0= -0.727$ and $D_0=-1.133$ yields a particle stuck and static at 
$(1,0.4)$ after the departure of the pulse. Note that the centre of the circular trajectory was initially at $(1,2)$. Figure 5 shows this result. 

\noindent In addition,  for certain chosen values of $\omega_B T$ we have the following scenarios.
When $\omega_B T = 2n \pi$ ($n=0,1,2...$), the radius of the circle does not change, though its
centre shifts. If, $\omega_B T = (2n+1)\pi$ ($n=0,1,2,...$), the radius changes
according to the chosen value of $\alpha$. For $\omega_B T = - 2 \alpha$,
the radius always decreases as long as $0<\alpha<\pi$.

\

\noindent {\sf Relative separation:} What happens when we consider a pair of identical charges starting out with different initial conditions? 
To analyse the trajectories, we make the following choices for the constants:

\centerline{ {\bf Charge 1:} $A_{01}$, $D_{01}$, $C_{01}$, $C'_{01}$\,\, ; \,\,
{\bf Charge 2:} $A_{02}$, $D_{02}$, $C_{02}$, $C'_{02}$.}

\noindent The radii of the circular orbits before the charges encounter the
pulse are $R_{i1}=\sqrt{A_{01}^2 + D_{01}^2}
$ and $R_{i2}= \sqrt{A_{02}^2 + D_{02}^2}$. Hence the difference in the
magnitudes of the radii and velocities are given as:

\centerline{ $\Delta R_i = \vert R_{i2}- R_{i1}\vert$ \,\, , \,\,$\Delta v_{i}= \vert v_{i2}-v_{i1} \vert = \omega_B \Delta R_i$.}

\begin{figure}[h]
\includegraphics[width=0.45\textwidth]{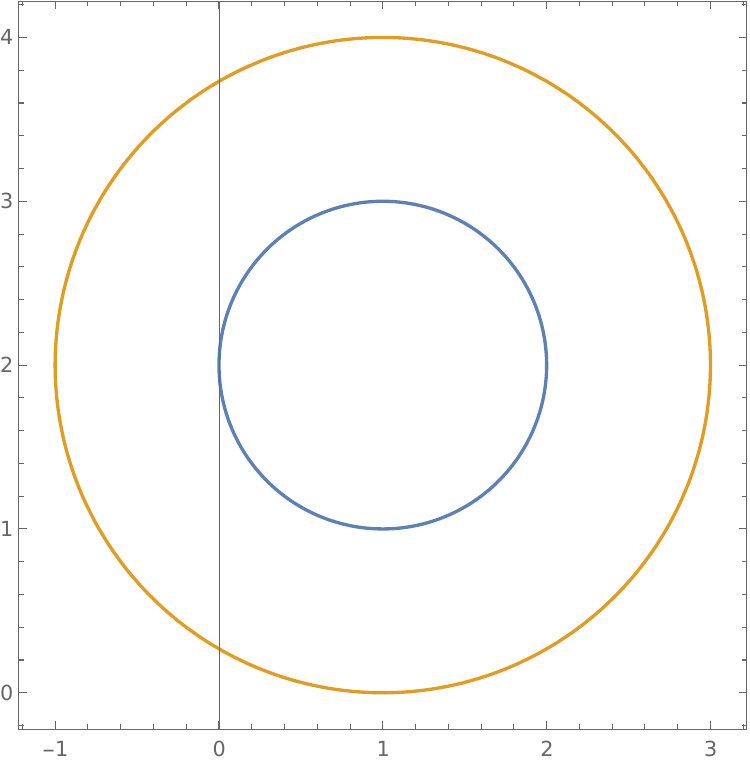}
\includegraphics[width=0.45\textwidth]{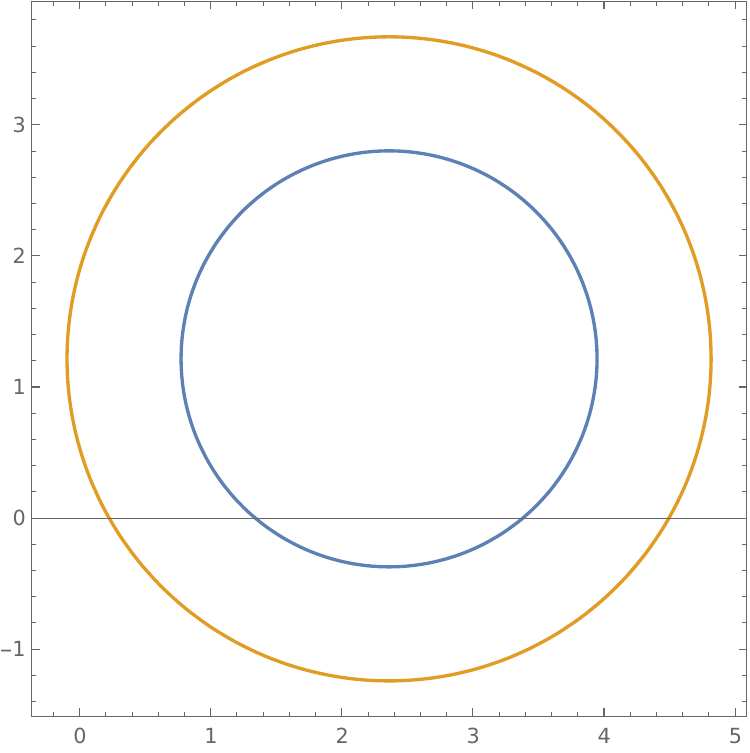}
\caption{The left figure shows the initial circular trajectories
(before the arrival of the pulse) of radii $R_{i1}$
(blue)
and $R_{i2}$ (yellow). The right figure shows the final circular trajectories
(after the pulse has departed)
of radii $R_{f1}$ (blue) and $R_{f2}$ (yellow). The chosen values are
$A_{01} = D_{01} = \frac{1}{\sqrt{2}}$, $A_{02}= D_{02} = \sqrt{2}$,
$C_{01} = C_{02} =1$, $C'_{01}=C'_{02}=2$,
$\omega_B=1$, $T=
\pi$, $\frac{E}{B} = 0.5$, $\alpha=\frac{\pi}{4}$. Note that $\Delta R_i \neq \Delta R_f$
and hence $\Delta v_i \neq \Delta v_f$. In the graphs $\Delta R_i = 1$, $\Delta R_f = 0.8219$. The concentric centre of the two particles is initially at $(1,2)$. After the pulse 
departs it shifts to $(2.11,0.89)$.}
\end{figure}

\noindent After the pulse has interacted and departed, we have different radii
and velocities. In particular, we have,
\begin{eqnarray}
R_{f1}^2= \left [ A_{01} +\frac{2 E}{\omega_B B }\sin \, \frac{\omega_B T}{2} \cos \left (\frac{\omega_B T}{2} +\alpha \right ) \right ]^2 + \nonumber \\ \left [ D_{01} +\frac{2 E}{\omega_B B }\sin \, \frac{\omega_B T}{2} \sin \left (\frac{\omega_B T}{2} +\alpha \right ) \right ]^2 .
\end{eqnarray}
\begin{eqnarray}
R_{f2}^2= \left [ A_{02} +\frac{2 E}{\omega_B B }\sin \, \frac{\omega_B T}{2} \cos \left (\frac{\omega_B T}{2} +\alpha \right ) \right ]^2 + \nonumber \\ \left [ D_{02} +\frac{2 E}{\omega_B B }\sin \, \frac{\omega_B T}{2} \sin \left (\frac{\omega_B T}{2} +\alpha \right ) \right ]^2.
\end{eqnarray}

\noindent Therefore,

\centerline{ $\Delta R_f= \vert R_{f2}- R_{f1}\vert$ \,\, , \,\,$\Delta v_{f}= \vert v_{f2}-v_{f1} \vert = \omega_B \Delta R_f$.}

\noindent From the expressions above it is clear that $\Delta R_i \neq \Delta R_f$
and hence $\Delta v_i \neq \Delta v_f$. 
Further, the centre also shifts. This is shown in Figure 6. which also
demonstrates a `permanent change' character of the {\bf memory-like effect}. Note that in articles on gravitational wave memory (eg. \cite{bragri},\cite{zhang1}, \cite{ic1}, \cite{ic4}, \cite{eanna1}, \cite{eanna2}) one usually
analyses the change in `relative separation' between pairs of test particles, which is qualitatively analogous to studying  geodesic deviation.
In our discussion above, we have, in a similar way, also shown the
presence of a net change in the relative separation. 

\

\noindent {\sf Experimental possibility:} It may be possible to see the above-analysed memory effect in a cyclotron.
Recall that one uses a radio-frequency alternating electric field (equal to
$\omega_B$ and spread over a
small region along the diameter of the dees) between the dees of
a cyclotron, in order to drive the
electron along a spiral path in the dees. In this process, the electron is also 
accelerated, which is indeed the purpose in a cyclotron. One will have to modify the
set-up a bit so that one has a homogeneous, short-duration electric field apart from the magnets which give the uniform magnetic field. It may then be possible to see the change in the circular trajectory, caused by the pulse -- i.e. see the change in radius, the shift of the centre and the change in the velocity. The practical
implementation of this idea in a possible experiment will surely require more
detailed thought which is left for the interested experimentalist to worry about.

\noindent In order to get a realistic picture, let us now find some numbers. We assume $m= 3.3 \times
{10}^{-27}$ kg (deuteron) , $q= 1.6 \times {10}^{-19}$ C and $B=1.5$ Tesla.
$\alpha$ is chosen to be equal to $2\pi$.
These values imply $\omega_B = 72.7 \times {10}^{6}$ rad/s which is in the
radio frequency range (around 12 MHz). We also assume $T=10$ nanoseconds and $E=50$ MegaVolts/m. The constants are chosen as $A_0 =0.30$ m, $D_0=0.40$ m,
$C_0=0.60$ m and $C'_0 =-0.10$ m. The chosen values yield the following:

{Initial trajectory:  Centre at $(0.60,-0.10)$, Radius $R_i= 0.50$ m
and $v_i = 0.12 $ c}.

{Final trajectory:  Centre at $(0.60,-0.43)$, Radius $R_f= 0.79$ m
and $v_f= 0.19$ c}.

\noindent From these numbers one may get a quantitative idea of the amount by which the centre may shift and how much the radius and velocity can change due to a short duration electric pulse. Note that all the velocities are non-relativistic in value.

\

\noindent {\sf Current source:} One may write down the current ${\bf J}$ which produces the
electric pulse using the Ampere-Maxwell law. The electric pulse is written as
\begin{eqnarray}
{\bf E} =  E \left [\Theta (t) - \Theta(t-T) \right ] \left (\cos \, \alpha \, {\bf \hat i} +
\sin \, \alpha \, {\bf \hat j} \right ).
\end{eqnarray}
This gives a  current ${\bf J}$ (in vacuum) equal to
\begin{eqnarray}
{\bf J} = \epsilon_0 E \left [ \delta (t) - \delta(t-T)\right ] \left (\cos \, \alpha \, {\bf \hat i} +
\sin \, \alpha \, {\bf \hat j} \right ).    
\end{eqnarray}
The current appears as a spike at $t=0$, creates the uniform electric field which is
sustained till $t=T$. At $t=T$, there is a current spike along the opposite direction 
which reduces the net electric field to zero. This is how the short duration pulse seems 
to act.

\section{\bf THE ROLE OF THE VECTOR POTENTIAL }

\noindent In the calculation presented above, we have seen a permanent change
in the trajectory of the charged particle after the pulse is switched off.
We have tried to suggest that this is indeed similar to `memory' as understood
in radiative contexts involving gravitational or electromagnetic waves. Here we
provide a somewhat stronger argument 
which, hopefully, will bring our results closer to what is usually called `memory'.

\noindent Firstly, as stated at the outset, one must note that here we do not have the standard `null' coordinate $u$ used in memory 
calculations, since we are in a non-relativistic regime. $u$ here is simply replaced by ordinary time  $t$. 

\noindent In our example in the previous sections we used an electric field pulse ${\bf E}$
and a uniform ${\bf B}$. The vector potential $\bf A ({\bf r}, t)$ which generates
the chosen $\bf E$ and $\bf B$ may be written as:
\begin{eqnarray}
{\bf A} = -\frac{1}{2} B \left (y {\bf \hat i} - x {\bf \hat j} \right ) +
\left (\cos \alpha {\bf \hat i} + \sin \alpha {\bf \hat j} \right ) F(t)
\end{eqnarray}
where,
\begin{eqnarray} 
F(t) = 0  \hspace{0.4in}  t \leq 0 \\
= -E \, t  \hspace{0.2in}  0\leq t \leq T \\
=  -E \, T  \hspace{0.4in}  t \geq T
 \end{eqnarray}
The first term in (30), which is purely spatial, generates the uniform ${\bf B}  = \nabla \times {\bf A}$ i.e. a constant magnetic field along the $z$ 
direction, present for all time. The second term in (30), which we will refer to as ${\bf A^{(t)}}$, is responsible for
the square pulse electric field, as evident from the defining relation ${\bf E} = - \frac{\partial {\bf A}}{\partial t} = -\frac{d {\bf A^{(t)}}}{dt}$.  It is worth noting here that the vector potential
is different in the regions $t<0$ and $t>T$ though the $\bf E$, $\bf B$ fields remain the same, before and after the duration of the pulse. Figure 7 (the blue curve) shows the temporal part of the vector potential as a function of time along with the electric field pulse (yellow curve).
The difference {$\Delta \bf A^{(t)}$} between the temporal parts of the vector potentials before and after the 
duration of the pulse is a non-zero constant vector,
given by,
\begin{eqnarray}
\Delta {\bf A^{(t)}} = - ET \left (cos\alpha \, {\bf \hat i} + \sin \alpha \, {\bf \hat j}\right )
\end{eqnarray}
It is this change which manifests itself in the differences in the
circular trajectories (shift of the centre, change in radius and velocity) and   is responsible for a memory-like effect.
\begin{figure}[h]
\includegraphics[width=0.6\textwidth]{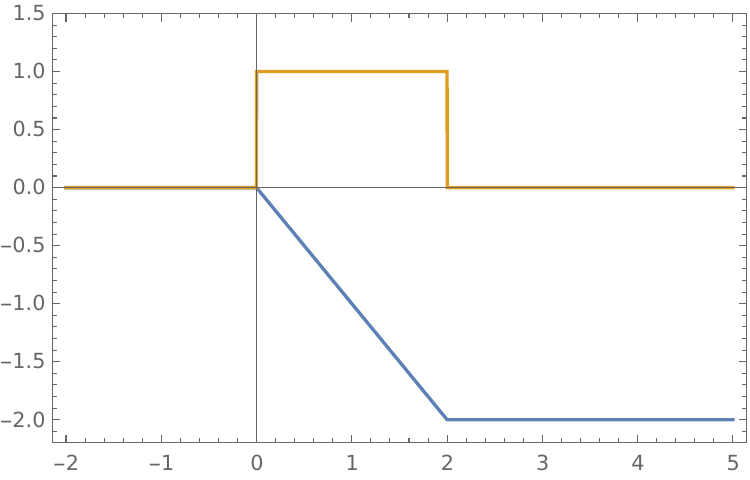}
\caption{The electric field pulse (yellow curve) for $\alpha=0$, $E=1$ unit and $T=2$ units. The temporal part of the vector potential (blue curve) for $\alpha=0$, $E=1$ unit and $T=2$ units.}
\end{figure}
\noindent More importantly,  we may relate the vector potential before and after the pulse
via a gauge transformation. Since ${\bf A_{before}}$ is related to ${\bf A_{after}}$ via ${\bf A_{after}} =  {\bf A_{before}} +{\nabla \Lambda}$, we obtain,
\begin{eqnarray}
\Lambda = -E T \left (x \cos \alpha + y\sin \alpha \right )
\end{eqnarray}
Hence, it should be clear that our memory-like effect has its roots in the
gauge-related change in the value (asymptotic values in the past and future) of the vector potential ${\bf A}$. 
Even though the change is a constant shift, it affects motion. 
The fact that such a gauge transformation between 
vector potentials in the asymptotic past and future is associated with memory has been noted by others too \cite{susskind, madwin2}. 

\noindent In the special case where the velocity kick is given by the time integral of the electric field vector only (i.e. no magnetic field or scalar potential are present) one can clearly see
that it is equal to the difference in the asymptotic values of the
vector potential. More precisely, we have the positions and velocities
in the three regions given as:

\noindent {\bf Region I ($t\leq0$):}
\begin{eqnarray}
    x^I = v_{x0} t +x_0 \hspace{0.2in};\hspace{0.2in} v_x^I =v_{x0}
    \hspace{0.2in};\hspace{0.2in}
    y^I = v_{y0} t +y_0 \hspace{0.2in};\hspace{0.2in} v_y^I =v_{y0}
\end{eqnarray}
 
\noindent {\bf Region II ($0\leq t\leq T$):}
\begin{eqnarray}
    x^{II} = x_0 +v_{x0} t + \left (\frac{qE\cos \,\alpha}{m}\right ) \frac{t^2}{2} \hspace{0.2in};\hspace{0.2in} v_x^{II} = v_{x0} + \left (\frac{qE\cos \,\alpha}{m}\right ) t \\
y^{II} = y_0 +v_{y0} t + \left (\frac{qE\sin \,\alpha}{m}\right ) \frac{t^2}{2} \hspace{0.2in};\hspace{0.2in} v_y^{II} = v_{y0} + \left (\frac{qE\sin \,\alpha}{m}\right ) t
\end{eqnarray}

\noindent {\bf Region III ($ t\geq T$):}

\begin{eqnarray}
    x^{III} = x_0 +v_{x0} t + \left (\frac{qE\cos \,\alpha}{m}\right ) T\,t -  \left (\frac{qE\cos \,\alpha}{m}\right ) \frac{T^2}{2} \hspace{0.12in};\hspace{0.12in} v_x^{III} = v_{x0} + \left (\frac{qE\cos \,\alpha}{m}\right )T \\
    y^{III} = y_0 +v_{y0} t + \left (\frac{qE\sin \,\alpha}{m}\right ) T\,t -  \left (\frac{qE\sin \,\alpha}{m}\right ) \frac{T^2}{2}
 \hspace{0.12in};\hspace{0.12in} v_y^{III} = v_{y0} + \left (\frac{qE\sin \,\alpha}{m}\right )T
\end{eqnarray}
It is easy to verify that $\Delta{\bf v} = -\frac{q}{m}\Delta {\bf A}= - \frac{q}{m} \nabla \Lambda$ ($\Delta {\bf v} = {\bf v}(t>T) - {\bf v}(t<0)$, $\Delta {\bf A} = {\bf A} (t>T) - {\bf A} (t<0)$),
using the expressions in (34), (35). One also notices that for a pair of test charges with different conditions at $t=0$, there is no change
in relative separation.

\noindent However, the kick is not directly given by just a
difference of asymptotic values of ${\bf A}$, when we have an ambient magnetic field (uniform or non-uniform). Integrating Eqns. (4), (5) from large negative to large positive values of $t$, one obtains,
\begin{eqnarray}
\Delta {\bf v} = -\frac{q}{m} {\Delta }{\bf A^{(t)}} +\omega_B \Delta {\bf \tilde r} 
\end{eqnarray}
where $\Delta {\bf v} = {\bf v}(t>T) - {\bf v}(t<0)$,
$\Delta {\bf A^{(t)}} = {\bf A^{(t)}}(t>T) - {\bf A^{(t)}}(t<0)$, ${\bf \tilde r} = y { \bf \hat i} - x{\bf \hat j}$ and $\Delta { \bf \tilde r} = {\bf \tilde r}(t>T) - { \bf \tilde r}(t<0)$. 
Equivalently, one may use the total vector potential ${\bf A}$
instead of $\bf A^{(t)}$ in (41), though one needs to keep in mind that
$\frac{d{\bf A}}{dt} =  \frac{\partial{\bf A}}{\partial t} + \left ({\bf v} \cdot \nabla\right ) {\bf A}  = \frac{\partial{\bf A}}{\partial t}-\frac{B}{2} \frac{d {\bf \tilde r}}{dt}$ and, therefore,
$\Delta {\bf A^{(t)}} = \Delta {\bf A} + \frac{B}{2} {\Delta {\bf \tilde r}}$. Thus, we may rewrite (41) as: $\Delta {\bf v} = -\frac{q}{m} {\Delta }{\bf A} +\frac{\omega_B}{2} \Delta {\bf \tilde r}$.

\noindent Using the expressions for the positions and velocities in 
regions I and III and the components of ${\bf A^{(t)}}$, one may easily verify (41) from both sides. Working explicitly with vector components, it is easy to see how the net difference in the vector potential plays a crucial role in ensuring the equality here. Eqn. (41), which visibly demonstrates the 
need and presence of the difference in the vector potential, is another generalization of the velocity kick formula in (1), when we have a uniform magnetic field perpendicular to the plane of motion. When $\omega_B=0$, it reduces to the case without a magnetic field. 
In addition, as shown earlier, when both the magnetic field and the electric pulse are present, there is indeed a change in relative separation.

\section{Conclusions}

\noindent In this brief article, we have shown how an effect somewhat similar to what is conventionally termed as electromagnetic memory, can arise in 
a rather simple scenario. The age-old problem of a charged particle moving on a
circular path due to a uniform magnetic field is modified by introducing a short-duration
electric pulse. This causes a `permanent change'--the radius can increase/decrease, 
the centre of the circle always shifts away and the velocity as well as its magnitude can increase or decrease. 
We also showed that for a pair of particles moving in the same ${\bf B}$ and ${\bf E}$
fields but with different conditions at a given time, the relative separation, velocity do change. The origins of this `permanent change', as explained in the previous section, can be traced to a net difference in the vector potential which, in turn, is linked to a gauge transformation.  This feature is analogous to memory as found using a gravitational wave pulse where, a change in relative separation has its origins (for asymptotically flat spacetimes) in diffeomorphisms of flat spacetime at asymptotic infinity. One may also distinguish between
pulses `with memory' or `without memory' for the given system, using the change in relative separation as a signature.
Finally, we have also suggested (with some explicit values) how our results may be seen in a cyclotron with necessary practical 
adjustments.

\noindent As stated in the Introduction, it is true that 
the example discussed here does not concern the usual `wave memory' caused by radiation at null infinity. It is, rather,
a completely non-relativistic scenario where the effect of a short duration pulse 
on a trajectory is explored. One may argue that this can indeed also happen in 
simple mechanical systems -- say, when a particle is acted upon by an impulsive force.
In particle mechanics though, the particle will actually have to be `struck' in order to
give an impulse. This is not the case when a charged particle moves in magnetic and electric fields--the interaction of the charge with the magnetic and electric field is not
a `contact' interaction and may be visualised as an `influence' over a short duration, 
which leads to a change which persists. Even though here we do not have a 
`radiative wave field' at null infinity, the
`time dependent pulse' can bring about a change akin to memory. Further, we have shown how our memory-like effect is related to the asymptotic (in time) values of the vector potentials which are gauge-related. 
A little search in the literature reveals that such
a pulse-induced memory is not uncommon in condensed matter (magnetic) systems as is well-known and well-studied in various articles (see eg. \cite{jacs}, \cite{acs} among others).

\noindent An obvious extension of this work is to try out different pulse shapes
--delta function, triangular, Gaussian, sech-squared are the simplest examples
--the last two being smooth functions. It may be difficult to do the smooth function
examples analytically, but all scenarios are surely tractable numerically. An important 
related question is about whether there is a pulse for which we can observe
a shape change in the trajectory--i.e. a circle going over to an ellipse or some other closed curve.
Such a model will be useful because it is, in some sense, perhaps closer to the idea of memory
as proposed and pursued in the context of gravitational wave physics \cite{ayansayan}. As a concluding remark we mention that it may be possible to extrapolate the basic ideas
presented here to the context of gravitoelectromagnetism \cite{gem1,gem2}
where, at a linearised level, one recovers the Maxwell and Lorentz force
equations with the  `electric' and `magnetic' fields born out of the Riemann tensor components of a background spacetime geometry. 

\noindent \section*{ACKNOWLEDGEMENTS}

\noindent The author thanks Indranil Chakraborty for his comments and suggestions
on the manuscript.
\noindent \section*{DATA AVAILABILITY STATEMENT}

\noindent No Data associated in the manuscript.

\end{document}